\documentstyle[11pt,aaspp4,flushrt]{article} 

\begin{document}
\title{The Rate of Turbulent Spherical Accretion} 
\author{Andrei Gruzinov}
\affil{Institute for Advanced Study, School of Natural Sciences, Princeton, NJ 08540}

\begin{abstract}
The rate of turbulent spherical accretion onto a compact object might be much smaller than the Bondi rate. It is suggested that the rate of accretion onto Sgr A$^*$ is much smaller than the Bondi rate.

\end{abstract}

\section{The rate of spherical accretion}
The rate of adiabatic inviscid spherical accretion onto an object of mass $M$ is given by the Bondi formula
\begin{equation}
\dot{M}_{\rm Bondi}\sim \rho c_sR_A^2,
\end{equation}
where $\rho$ is the density of the accreting gas, $c_s$ is the speed of sound, and $R_A$ is the accretion radius
\begin{equation}
R_A\sim GMc_s^{-2}.
\end{equation}
We show that turbulent heat conduction reduces the rate of accretion onto a compact object 
\begin{equation}
\dot{M}\sim\dot{M}_{\rm Bondi}\left( {R_S\over R_A}\right) ^{\alpha },
\end{equation}
where $R_S\ll R_A$ is the radius of the compact object, and $\alpha$ is a model-dependent positive parameter. This formula is derived in the next section. Turbulent viscosities are not considered in this work, but they should also slow down the accretion. Turbulent heat conduction hinders spherical accretion because it saves part of the heat from being accreted onto the object. The accretion flow becomes hotter, and hot gas accretes slower. 

To illustrate the astrophysical applications of this result consider the case of Sgr A$^*$, which is believed to be a $2.5\times 10^6M_{\odot }$ black hole located at the center of our Galaxy; its accretion luminosity might be $L\sim 10^{37}{\rm erg}/{\rm s}$ (Genzel et al 1994). When stellar winds in the vicinity of the galactic center collide and shock, they produce a gas of $c_s\sim 1000{\rm km}/{\rm s}$ and $\rho \sim 10^{-20}{\rm g}/{\rm cm}^3$ (Coker \& Melia 1997). The Bondi accretion rate of this gas is $\dot{M}_{\rm Bondi}\sim 10^{21} {\rm g}/{\rm s}$. Then $\dot{M}_{\rm Bondi}c^2\sim 10^{42} {\rm erg}/{\rm s}$ is five orders of magnitude higher than the actual luminosity. We (Gruzinov 1998, Quataert \& Gruzinov 1998) and others (Meszaros 1975, Blandford 1998) have argued that such low radiative efficiencies are unrealistic. It is then natural to assume that Sgr A$^*$ accretes at a much smaller rate (Blandford \& Begelman 1998). If we use the estimate (3), and arbitrarily assume $\alpha = 0.4$, the estimated accretion rate is reduced by a factor of 100. The radiative efficiency is then $0.1\% $. This radiative efficiency is still small, but it might be reasonable because electrons might be heated much less than ions, and even hot electrons might radiate inefficiently when the plasma is rarefied. 

\section{A solvable model}  
Here we derive equation (3). We assume a spherically symmetrical gas inflow. The flow is taken to be inviscid but non-adiabatic, i.e. we include turbulent thermal conduction and neglect turbulent viscosities. This artificial assumption is made to simplify the model. No real calculation is possible anyway, since we do not understand the nature of MHD turbulence in the accretion flow. Our sole purpose is to demonstrate that the rate of turbulent spherical accretion might be much smaller than the Bondi rate. 

The gas density $\rho$, the inflow (positive) velocity $v$, and temperature $T$ depend on the radial coordinate $r$ only, and satisfy the stationary equations of continuity, Euler's, and the thermal conduction
\begin{equation} 
\rho vr^2=J,
\end{equation}
\begin{equation} 
vv'=-\rho ^{-1}(\rho T)'-r^{-2},
\end{equation}
\begin{equation} 
\rho Tv\left( {\rho '\over \rho }-{3\over 2}{T'\over T}\right) = r^{-2}(\kappa r^2T')'.
\end{equation}
Here $J\equiv \dot{M}/4\pi $, the prime denotes the radial derivative, $GM=1$, $m_p=1$. We assume that the thermal conductivity $\kappa$ is given by the ``Shakura-Sunayev'' formula
\begin{equation} 
\kappa =\alpha \rho rv,
\end{equation}
where $\alpha \sim 1$ is a dimensionless positive constant. 

The boundary condition at the surface of the compact object is unknown, but it is irrelevant. For concreteness, assume $T+\beta rT'=0$ at $r=R_S$, where $\beta$ is a dimensionless constant. At $r=\infty$, we assume $T=1$, so that $R_A\sim 1$. The boundary condition for $\rho$ is also irrelevant, because the system of equations is invariant under $\rho \rightarrow \lambda \rho$, $J \rightarrow \lambda J$. 

Our aim is to find a maximal possible value of $\dot{m}\equiv J/\rho (\infty )$, for which a smooth solution of the system exists. The ratio $\dot{M}/\dot{M}_{\rm Bondi}$ is equal to this maximal value. Obviously $\dot{m}\sim 1$ if $\alpha \sim 1$ and $R_S\sim 1$. We need to find out how $\dot{m}$ depends on $R_S$ when $R_S\ll 1$. 

Scale out $J$ by $\rho \rightarrow J\rho$, plug (4) and (7) into (6):
\begin{equation} 
\rho vr^2=1,
\end{equation}
\begin{equation} 
vv'=-\rho ^{-1}(\rho T)'-r^{-2},
\end{equation}
\begin{equation} 
T\left( {\rho '\over \rho }-{3\over 2}{T'\over T}\right) = \alpha (rT')'.
\end{equation}
Using (10), integrate (9) and obtain a system of two first-order differential equations
\begin{equation} 
{1\over 2} {1\over r^4\rho ^2}+{5\over 2}T+\alpha rT'={1\over r}+{5\over 2},
\end{equation}
\begin{equation} 
{\rho '\over \rho}\left( T-{1\over r^4\rho ^2}\right) =-{1\over r^2}+{2\over r^5\rho ^2}-T'.
\end{equation}
We need to find how the  maximal possible value of $\dot{m}\equiv 1/\rho (\infty )$ (for which a smooth solution of the system exists) depends on $R_S\ll 1$. For small $r$, we can neglect $5/2$ in the right hand side of (11). Then, denoting $\tau=rT$, $f=(r^3\rho ^2)^{-1}$, $x=\log r$, we obtain 
\begin{equation} 
2\alpha {d\tau \over dx}=2-(5-2\alpha)\tau -f,
\end{equation}
\begin{equation} 
{df \over dx}=f{2+2d\tau /dx-5\tau -f\over \tau -f}.
\end{equation}
This system has a stable equilibrium point $\tau =2(5-2\alpha)^{-1}$, $f=0$. In the vicinity of this point, 
\begin{equation} 
{df \over dx}=-2\alpha f.
\end{equation} 
The system (13), (14) is an accurate approximation only for $x<0$, that is for an x-duration $\delta x=\log (1/R_S)$. The quantity $f$ decreases according to (15) for an x-duration $\delta x=\log (1/R_S)-{\rm const}$. Therefore the final density is proportional to $(R_S)^{-\alpha }$. Since $\dot{m}\equiv 1/\rho (\infty )$, the scaling law (3) is proven. To check the answer we integrated (11), (12) numerically.

\acknowledgements This work was supported by NSF PHY-9513835.

\end{document}